\documentclass[twocolumn,english,conference]{IEEEtran}
\usepackage{lmodern}
\usepackage[T1]{fontenc}
\usepackage{babel}
\usepackage{units}
\usepackage{amsthm}
\usepackage{amsmath}
\usepackage{amssymb}
\usepackage[all]{xy}
\usepackage[unicode=true,
 bookmarks=true,bookmarksnumbered=true,bookmarksopen=true,bookmarksopenlevel=1,
 breaklinks=false,pdfborder={0 0 0},backref=false,colorlinks=false]
 {hyperref}
\hypersetup{pdftitle={Your Title},
 pdfauthor={Your Name},
 pdfpagelayout=OneColumn, pdfnewwindow=true, pdfstartview=XYZ, plainpages=false}

\makeatletter
\theoremstyle{plain}
\newtheorem{thm}{\protect\theoremname}
\theoremstyle{definition}
\newtheorem{defn}[thm]{\protect\definitionname}
\theoremstyle{plain}
\newtheorem{prop}[thm]{\protect\propositionname}
\theoremstyle{plain}
\newtheorem{cor}[thm]{\protect\corollaryname}
\theoremstyle{plain}
\newtheorem{lem}[thm]{\protect\lemmaname}
\theoremstyle{plain}
\newtheorem{conjecture}[thm]{\protect\conjecturename}
\theoremstyle{definition}
\newtheorem{example}[thm]{\protect\examplename}

\usepackage[caption=false,font=footnotesize]{subfig}

\usepackage{amsfonts}
\usepackage{pgf}
\usepackage[all]{xy}

\newcommand{\xyR}[1]{
  \xydef@\xymatrixrowsep@{#1}}
\newcommand{\xyC}[1]{
  \xydef@\xymatrixcolsep@{#1}}

\newdir{|>}{!/4.5pt/@{|}*:(1,-.2)@^{>}*:(1,+.2)@_{>}}

\let\myTOC\tableofcontents
\renewcommand\tableofcontents{%
  \pdfbookmark[1]{\contentsname}{}
  \myTOC }

\def\LyX{\texorpdfstring{%
  L\kern-.1667em\lower.25em\hbox{Y}\kern-.125emX\@}
  {LyX}}

\usepackage{ifpdf}
\ifpdf

\IfFileExists{lmodern.sty}
 {\usepackage{lmodern}}{}

\fi 

\makeatother

\providecommand{\conjecturename}{Conjecture}
\providecommand{\corollaryname}{Corollary}
\providecommand{\definitionname}{Definition}
\providecommand{\examplename}{Example}
\providecommand{\lemmaname}{Lemma}
\providecommand{\propositionname}{Proposition}
\providecommand{\theoremname}{Theorem}

\begin{document}

\title{Lattices with non-Shannon Inequalities}

\author{\IEEEauthorblockN{Peter~Harremo{\"e}s}\IEEEauthorblockA{Copenhagen Business College\\
Copenhagen\\
Denmark\\
Email: harremoes@ieee.org}}
\maketitle
\begin{abstract}
We study the existence or absence of non-Shannon inequalities for
variables that are related by functional dependencies. Although the
power-set on four variables is the smallest Boolean lattice with non-Shannon
inequalities there exist lattices with many more variables without
non-Shannon inequalities. We search for conditions that ensures that
no non-Shannon inequalities exist. It is demonstrated that 3-dimensional
distributive lattices cannot have non-Shannon inequalities and planar
modular lattices cannot have non-Shannon inequalities. The existence
of non-Shannon inequalities is related to the question of whether
a lattice is isomorphic to a lattice of subgroups of a group.\end{abstract}

\IEEEpeerreviewmaketitle{}

\section{Introduction}

The existence of non-Shannon inequalities have got a lot of attention
since the first inequality was discovered by Z. Zhang and R. W. Yeung
\cite{Zhang1988}. The basic observation is that any four random variables
$A,$ $B,$ $C,$ and $D$ satisfy the following inequality
\begin{multline}
2I\left(C;D\right)\leq\\
I\left(A;B\right)+I\left(A;C\uplus D\right)+3I\left(C;D\mid A\right)+I\left(C;D\mid B\right)\label{eq:ZhangYeung}
\end{multline}
where $I\left(A,B\right)$ denotes the mutual information between
$A$ and $B$ and $I\left(C;D\mid B\right)$ denotes the conditional
mutual information between $C$ and $D$ given $B.$ Finally, $C\uplus D$
here denotes the random variable that takes values of the form $\left(c,d\right)$
where $c=C$ and $d=D.$ The inequality is non-Shannon in the sense
that it cannot be deduced from inequalities of the form 
\begin{eqnarray*}
H\left(X\uplus Y\right) & \geq & H\left(X\right)\\
I\left(X;Y\mid Z\right) & \geq & 0.
\end{eqnarray*}
Using that $I\left(X;Y\right)=H\left(X\right)+H\left(Y\right)$ and
\begin{multline*}
I\left(X;Y\mid Z\right)=\\
H\left(X\uplus Z\right)+H\left(Y\uplus Z\right)-H\left(X\uplus Y\uplus Z\right)-H\left(Z\right)
\end{multline*}
the last inequality can be rewritten as
\[
H\left(X\uplus Z\right)+H\left(Y\uplus Z\right)\leq H\left(X\uplus Y\uplus Z\right)+H\left(Z\right),
\]
which we will call the sub-modular inequality. Therefore the Shannon
inequalities are the ones that can be deduced from using that entropy
is non-negative, increasing and submodular. Later it was shown by
F. Matus \cite{Matus2007} that for four variables there exists infinitely
many non-Shannon inequalities. It is easy to show that any inequality
involving only three variables rather than four can be deduced from
Shannon's inequalities. Now the power set of for variables is a Boolean
algebra with 16 elements and any smaller Boolean algebra corresponds
to smaller number of variables, so in a trivial sense the Boolean
algebra with 16 elements is the smallest Boolean algebra for which
there exists non-Shannon inequalities. 

In the literature on non-Shannon inequalities all inequalities are
expressed in terms of sets of variables and their joins. Another way
to formulate this is that the inequalities are stated for the free
$\uplus$-semilattice generated by a finite number of variables. In
this paper we will also consider intersection of variables. We note
that for sets of variables we have the inequality 
\[
I\left(X;Y\mid Z\right)\geq H\left(X\cap Y\mid Z\right).
\]
This inequality have even inspired some authors to see the notation
$I\left(\cdot\wedge\cdot\right)$ to denote mutual information.

Although non-Shannon inequalities have been known for more than a
decade they have found remarkable few applications compared with Shannon's
inequalities. One of the reasons for this is that there exists much
larger lattices that a Boolean algebra with 16 elements. The simplest
example is are the Markov chains. 
\[
X_{1}\to X_{2}\to X_{3}\to\dots\to X_{n}
\]
where $X_{1}$ determines $X_{2}$ which determines $X_{3}$ etc.
For such a chain one has
\[
H\left(X_{1}\right)\geq H\left(X_{2}\right)\geq H\left(X_{3}\right)\geq\dots\geq H\left(X_{n}\right)\geq0.
\]
 These inequalities are all instances of monotonicity of the entropy
function, and it is quite clear that these inequalities are sufficient
in the sense that for any sequence of values that satisfies these
inequalities there exists random variables related by a deterministic
Markov chain with these values as entropies.

In this paper we look at entropy inequalities for random variables
that are related by functional dependencies. Functional dependencies
gives an ordering of variables into a lattice. Such functional dependence
lattices have many applications in information theory, but in this
short note we will focus on the question how one can detect whether
a lattice of functionally related variables can non-Shannon inequalities.
In particular we are interested in determination of the ``smallest''
lattice with non-Shannon inequalities. Here we should note that there
are several ways of measuring the size of a lattice, and also note
that in order to achieve interesting results have have to restrict
our attention to special classes of lattices.

Non-Shannon inequalities have been studied using matroid theory but
matroids are equivalent to atomistic semimodular lattices. For the
study of non-Shannon inequalities it is more natural to look at general
lattices rather than matroids because many important applications
involve lattices that are not atomistic or not semimodular. For instance
the deterministic Markov chain gives a lattice that is not atomistic.
It is known that a function is entropic if and only if it can (approximately)
equal to the logarithm of the index of a subgroup in a group. The
lattice of subgroups of a Abelian group is modular and atomistic and
can be described by matroid theory. A switch from matroids to lattices
corresponds to a switch from Abelian groups to more general groups.

\section{Lattices of functional dependence\label{sec:Lattices-of-functional}}

Many problems in information theory and cryptography can be formulated
in terms functional dependencies. For instance one might be interested
in giving each member of a group part of a password in such a way
that no single person can recover the whole password but any two members
are able to recover the password. Here the password should be a function
of the variables known by any two members but not a function of a
variable hold by any single member. In this section we shall briefly
describe functional dependencies and their relation to lattice theory.
The relation between functional dependence and lattices has previously
been studied \cite{Demetrovics1989,Demetrovics1992,Levene1995,Harremoes2011g}.
The relation between functional dependencies and Bayesian networks
is described in \cite{Harremoes2015}. 

Inspired by Armstrong's theory of relational databases \cite{Armstrong1974}
we say that a relation $\to$ in a lattice $L$ satisfies \emph{Armstrong's
axioms} if it satisfies the following properties.

\textbf{Transitivity} If $X\to Y$ and $Y\to Z$, then $X\to Z.$ 

\textbf{Reflexivity} If $X\geq Y$, then $X\to Y.$

\textbf{Augmentation} If $X\to Y$, then $X\vee Z\to Y\vee Z.$

In a database $X\to Y$ should mean that there exists a function such
that $Y=f\left(X\right)$ obviously satisfies these inference rules
so as an axiomatic system it is sound. Armstrong proved that these
axioms form a complete set of inference rules . That means that if
a set $A$ of functional dependencies is given and a certain functional
dependence $x\to y$ holds in any database where all the functional
dependencies in $A$ hold then $x\to y$ holds in that database. Therefore
for any functional dependence $x\to y$ that cannot be deduced using
Armstrong's axioms the exist a database where the functional dependence
is violated \cite{Ullman1989,Levene1999}. As a consequence there
exists a database where a functional dependence holds if and only
if it can be deduced from Armstrong's axioms. A lattice element $X$
is said to be closed if $X\to Y$ implies that $X\geq Y.$ The smallest
lattice element greater than $X$ will be denoted $cl\left(X\right).$
\begin{thm}
\label{thm:completeArmstrong}The set of closed elements form a lattice.
For any finite lattice there exist a set of related variables such
that the elements of the lattice corresponds to closed sets under
functional dependence. 
\end{thm}
According to the theorem any lattice can be considered as a closed
set of variables under some functional dependence relation where $X\to Y$
if and only if $X\supseteq Y.$ On the set of closed sets the meet
operation is given by $X\cap Y$ and the join operation is given by
$X\uplus Y=cl\left(X\cup Y\right).$ We observe that the set of closed
sets is a subset of the original lattice that is closed under intersection.
Such a subset is called a $\cap$-semilattice or a closure system.
Any closure system defines a relation that satisfies Armstrong's axioms.

On a lattice \emph{submodularity} of a function $h$ is defined via
the inequality $h\left(X\right)+h\left(Y\right)\geq h\left(X\uplus Y\right)+h\left(X\cap Y\right)$.
A \emph{polymatroid function} on a lattice can then be defined as
a function that is non-negative, increasing and sub-modular. The relation
$h\left(X\uplus Z\right)+h\left(Y\uplus Z\right)=h\left(X\uplus Y\uplus Z\right)+h\left(Z\right)$
defines a relation denoted $\left(X\bot Y\mid Z\right)$ that satisfies
the properties:

\textbf{Existence} $\left(X\bot Y\mid X\right).$

\textbf{Symmetry} $\left(X\bot Y\mid W\right)$ if and only if $\left(Y\bot X\mid W\right).$

\textbf{Decomposition} If $\left(X\bot Y\uplus Z\mid W\right)$ then
$\left(X\bot Z\mid W\right).$

\textbf{Contraction} $\left(X\bot Z\mid W\right)$ and $\left(X\bot Y\mid Z\uplus W\right)$
implies $\left(X\bot Y\uplus Z\mid W\right).$ 

\textbf{Weak union} $\left(X\bot Y\uplus Z\mid W\right)$ implies
$\left(X\bot Y\mid Z\uplus W\right).$

We say that a relation that satisfies these properties is \emph{semi-graphoid}.
Note that we allow the elements to have non-empty intersection. Note
also that the existence property is normally not included in the list
of semi-graphoid properties. If $\left(B\bot B\mid A\right)$ we write
$A\supseteq_{\bot}B$. If $h$ denotes the Shannon entropy $H$ then
$A\supseteq_{\bot}B$ simply means that $H\left(B\mid A\right)=0$
or equivalently that $B$ is almost surely a function of $A.$
\begin{thm}
\label{thm:meetsemilattice}If $\left(L,\cap,\uplus\right)$ is a
lattice with a semi-graphoid relation $\left(\cdot\bot\cdot\mid\cdot\right)$
then the relation $\supseteq_{\bot}$ satisfies Armstrong's axioms.
The relation $\left(\cdot\bot\cdot\mid\cdot\right)$ restricted to
the lattice of closed lattice elements is semi-graphoid. If the semi-graphoid
relation $\left(\cdot\bot\cdot\mid\cdot\right)$ is given by a polymatroid
function $h$ then $h$ is also polymatroid if it is restricted the
lattice of closed elements.
\end{thm}

\section{Entropy in functional dependence lattices}
\begin{defn}
A polymatroid function $h$ on a lattice $L$ is said to be entropic
if there exists a function $f$ form $L$ into a set of random variables
such that $h\left(x\right)=H\left(f\left(x\right)\right)$ for any
element in the lattice. 
\end{defn}
Let $L$ denote a lattice and let $\Gamma\left(L\right)$ denote the
set of polymatroid functions on $L.$ Let $\Gamma^{*}\left(L\right)$
denote the set of entropic functions on $L$ and let $\bar{\Gamma}^{*}\left(L\right)$
denote the closure of this set. 
\begin{defn}
A lattice is said to be a \emph{Shannon lattice} if any polymatroid
function can be realized approximately by random variables, i.e. $\Gamma\left(L\right)=\bar{\Gamma}^{*}\left(L\right).$
\end{defn}
Both $\Gamma\left(L\right)$ and $\bar{\Gamma}^{*}\left(L\right)$
are polyhedral sets and often we may normalize the polymatroid functions
by requiring that the value at the maximal element is 1. One may then
check whether a lattice is a Shannon lattice by checking that the
extreme polymatroid functions are entropic.

From the definition we immediately get the following result. 
\begin{prop}
\label{prop:meetsemilattice}If $L$ is a Shannon lattice and $M$
is a subset that is a $\cap$-semilattice then $M$ is a Shannon lattice.
In particular all sub-lattices of a Shannon lattice are Shannon lattices.
\end{prop}
With these results at hand we can start hunting non-Shannon lattices.
We take a lattice that may or may not be a Shannon lattice. We find
the extreme polymatroid functions and for each extreme point we determine
the lattice of closed elements using Theorem \ref{thm:meetsemilattice}.
Each of these lattices of closed sets have a much simpler structure
than the original lattice and the goal is now to check if these lattices
are Shannon lattices or not. It turns out that there are quite few
of these reduced lattices and they could be considered as the building
blocks for larger lattices. 

The simplest lattice just has just two elements. The only normalized
polymatroid function takes the values zero and one. It is obviously
entropic.

We recall that an element $i$ is $\uplus$-irreducible if $i=x\uplus y$
implies that $i=x$ or $i=y$. An $\cap$-irreducible element is defined
similarly. An element is double irreducible if it is both $\uplus$-irreducible
and $\cap$-irreducible. The lattice denoted $M_{n}$ is a modular
lattice with a smallest element, a largest element and $n-2$ double
irreducible elements arranged in-between. 

\begin{figure}[tbh]
\[
\xymatrix{ &  & *+[o][F-]{}\\
*+[o][F-]{}\ar@{-}[urr] & *+[o][F-]{}\ar@{-}[ur] & *+[o][F-]{}\ar@{-}[u] & *+[o][F-]{}\ar@{-}[ul] & *+[o][F-]{}\ar@{-}[ull]\\
 &  & *+[o][F-]{}\ar@{-}[ull]\ar@{-}[urr]\ar@{-}[ul]\ar@{-}[ur]\ar@{-}[u]
}
\]
\protect\caption{Hasse diagram of the lattice $M_{7}$.}
\end{figure}
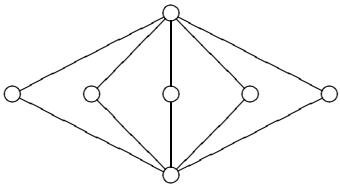

\begin{thm}
\label{thm:M_n}For any $n$ the lattice $M_{n}$ is a Shannon lattice.\end{thm}
\begin{IEEEproof}
The proof is essentially the same as the solution to the cryptographic
problem stated in the beginning of Section \ref{sec:Lattices-of-functional}.
The idea is that one should look for groups with a subgroup lattice
$M_{n}$ and then check that the subgroups of such group are actually
have the right cardinality.\end{IEEEproof}
\begin{cor}
\label{cor:trevaerdier}Any polymatroid function that only takes the
values $0,\nicefrac{1}{2}$, and $1$ is entropic.\end{cor}
\begin{IEEEproof}
Assume that the polymatroid function $h$ only takes the values $0,\nicefrac{1}{2},$
and $1$. Then $h$ defines a semi-graphoid relation and the closed
elements form a lattice isomorphic to $M_{n}$ for some integer $n$.
The function $h$ is entropic on $M_{n}$ so $h$ is also entropic
on the original lattice. 
\end{IEEEproof}
\begin{figure}[tbh]
\[
\xymatrix{ &  & *+[o][F-]{1}\\
*+[o][F-]{\nicefrac{3}{4}}\ar@{-}[urr] & *+[o][F-]{\nicefrac{3}{4}}\ar@{-}[ur] & *+[o][F-]{\nicefrac{3}{4}}\ar@{-}[u] & *+[o][F-]{\nicefrac{3}{4}}\ar@{-}[ul] & *+[o][F-]{\nicefrac{3}{4}}\ar@{-}[ull]\\
*+[o][F-]{\nicefrac{1}{2}}\ar@{-}[u]\ar@{-}[ur] & *+[o][F-]{\nicefrac{1}{2}}\ar@{-}[ul]\ar@{-}[ur]\ar@{-}[urr] &  & *+[o][F-]{\nicefrac{1}{2}}\ar@{-}[ull]\ar@{-}[ul]\ar@{-}[ur] & *+[o][F-]{\nicefrac{1}{2}}\ar@{-}[ul]\ar@{-}[u]\\
 &  & *+[o][F-]{0}\ar@{-}[ull]\ar@{-}[urr]\ar@{-}[ul]\ar@{-}[ur]
}
\]
\protect\caption{Lattice with a non-entropic polymatroid function. \label{fig:LLD}}
\end{figure}
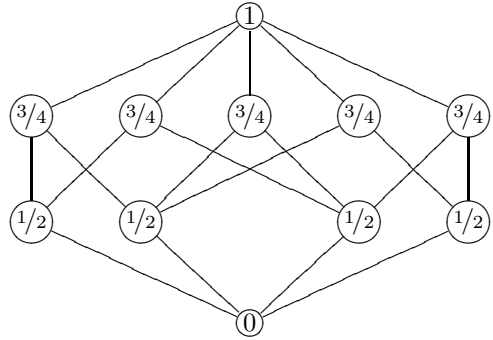
The Boolean lattice with four atoms is the smallest non-Shannon Boolean
algebra. Nevertheless there are smaller non-Shannon lattices. Figure
\ref{fig:LLD} illustrates a lattice with just 11 elements that violates
Inequality \ref{eq:ZhangYeung}. This corresponds to the fact that
the lattice in Figure \ref{fig:LLD} is not equivalent to a lattice
of subgroups of a finite group. The lattices that are equivalent to
lattices of subgroups of finite groups have been characterized \cite{Schmidt1994},
but the characterization is too complicated to describe in this short
note. Using the ideas from \cite{Matus2007} one can prove that the
lattice in Figure \ref{fig:LLD} has infinitely many non-Shannon inequalities.
Note that this lattice is atomistic but not semimodular, but it is
lower locally distributive. Any semimodular lattice that contains
the lattice in \ref{fig:LLD} as a $\cap$-semilattice also contains
a power set on four points as a $\cap$-semilattice. The following
lemma gives a considerable reduction in the number of inequalities
that one has to consider in search for extreme polymatroid functions.
\begin{lem}
\label{lem:increasing}If $h$ is submodular and increasing on $\cap$-irreducible
elements then $h$ is increasing.\end{lem}
\begin{thm}
\label{thm:minLLD}The lattice in figure \ref{fig:LLD} is the lower
locally distributive non-Shannon lattice with fewest elements.\end{thm}
\begin{IEEEproof}
There exists a nice presentation of lower locally distributive lattices
\cite{Behrendt1991} (In this paper the author works with the dual
lattices). With this representation one it is relatively simple to
create a list of all lower locally distributive lattices with 11 elements
or fewer. Each lattice has finitely many extreme polymatroid functions.
These can be found using the R program with package rcdd. Each of
these extreme polymatroid functions in each of these lattices has
been checked to be entropic.
\end{IEEEproof}
One may ask if there exists a lattice with fewer points than 11 that
is non-Shannon. 
\begin{thm}
Any lattice with 7 or fewer elements is a Shannon lattice.\end{thm}
\begin{IEEEproof}
Up to isomorphism there only exist finitely many lattices with 7 or
fewer elements. Each lattice has finitely many extreme polymatroid
functions. These can be found using the R program with package rcdd.
Each of these extreme polymatroid functions in each of these lattices
has been checked to be entropic.
\end{IEEEproof}

\section{Ingleton inequalities}

The polymatroid function on Figure \ref{fig:LLD} does not only violate
some non-Shannon inequalities, but it also violates an Ingleton inequality
\cite{Guille2011}. The Ingleton inequalities are inequities of the
form
\begin{multline*}
h\left(C\right)+h\left(D\right)+h\left(A\uplus C\uplus D\right)+h\left(B\uplus C\uplus D\right)+h\left(A\uplus B\right)\\
\leq\\
h\left(C\uplus D\right)+h\left(C\uplus A\right)+h\left(C\uplus B\right)+h\left(D\uplus A\right)+h\left(\uplus B\right).
\end{multline*}
A more instructive way of formulating the Ingleton inequalities is
in terms of conditional mutual information.
\begin{multline*}
I\left(X;Y\mid Z\right)\leq\\
I\left(X;Y\mid Z\uplus V\right)+I\left(X;Y\mid Z\uplus W\right)+I\left(V;W\mid Z\right).
\end{multline*}
 The Ingleton inequalities are satisfied for rank functions of representable
matroid. In particular all entropic functions that can be described
by Abelian groups satisfy the Ingleton inequalities. If a polymatroid
on a lattice satisfies the Ingleton inequality the associated semi-graphoid
relation satisfies the following property.

\textbf{Strong contraction} If $\left(X\bot Y\mid Z\uplus V\right)$
and $\left(X\bot Y\mid Z\uplus V\right)$ and $\left(V\bot W\mid Z\right)$
then $\left(X\bot Y\mid Z\right).$

Like the Ingleton inequality strong contraction does not hold for
all entropic polymatroid functions, but it does hold for most graphical
models of independence like Bayesian networks. Recently it was demonstrated
that strong contraction is essential for giving a lattice characterization
of an certain system of inference rules for conditional independence
\cite{Niepert2013}. In \cite{Niepert2013} strong contraction was
used in conjunction with the following property.

\textbf{Strong union} If $\left(X\bot Y\mid Z\right)$ then $\left(X\bot Y\mid Z\uplus W\right)$.

Strong union is a quite restrictive condition, but it does hold for
Markov chains and other Markov networks. The entropy inequality corresponding
to strong union is 
\[
I\left(X;Y\mid Z\right)\leq I\left(X;Y\mid Z\uplus W\right).
\]
If a polymatroid function satisfies the strong union inequality we
get a significant reduction in the complexity of the problem. 

Computer experiments support the following conjecture.
\begin{conjecture}
If a polymatroid function on a lattice satisfies the Ingleton inequalities
and the strong union inequalities then the function is entropic.
\end{conjecture}
It is worth noting that in \cite{Niepert2013} the authors use a lattice
technique that is slightly different from the one developed in the
present paper.

\section{Distributive and modular lattices}

The power-set of four variables is a distributive lattice so one may
ask if there exists any distributive lattice with non-Shannon inequalities
without this Boolean lattice as sub-lattice. We recall that a lattice
is said to be \emph{modular} if $a\subseteq b$ implies that 
\[
a\uplus\left(x\cap b\right)=\left(a\uplus x\right)\cap b
\]
for any lattice element $x.$ For modular lattices the following lemma
gives a considerable reduction in the number of inequalities that
one has to consider in the search for extreme points.
\begin{lem}
\label{lem:Modulaer}Let $L$ be a modular lattice with a function
$h$ that is submodular on any sub-lattice with elements $a,b,a\cap b$
and $a\uplus b$ where $a\cap b$ is covered by $a$ and b. Then the
function $h$ is submodular on $L.$ 
\end{lem}
For a distributive lattice the order dimension equals the maximal
number of $\uplus$-irreducible elements (or maximal number of $\cap$-irreducible
elements) needed in a decomposition of an element in the lattice.
Distributive lattices may also be represented as ideals in partially
ordered sets and the order dimension is also equal to the maximal
anti-chain in the partially ordered set used in such a representation.
\begin{thm}[\cite{Dilworth1950}]
Let L be a distributive lattice. Then L can be embedded as a sub-lattice
into the n-th power of a chain if and only if it has order dimension
at most $n.$
\end{thm}
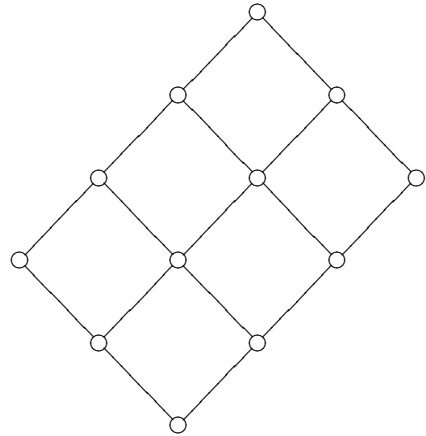
\begin{figure}[tbh]
\[
\xymatrix{ &  &  & *+[o][F-]{}\\
 &  & *+[o][F-]{}\ar@{-}[ur] &  & *+[o][F-]{}\ar@{-}[ul]\\
 & *+[o][F-]{}\ar@{-}[ur]\ar@{-}[dl] &  & *+[o][F-]{}\ar@{-}[ur]\ar@{-}[ul] &  & *+[o][F-]{}\ar@{-}[ul]\\
*+[o][F-]{} &  & *+[o][F-]{}\ar@{-}[ur]\ar@{-}[ul] &  & *+[o][F-]{}\ar@{-}[ur]\ar@{-}[ul]\\
 & *+[o][F-]{}\ar@{-}[ur]\ar@{-}[ul] &  & *+[o][F-]{}\ar@{-}[ur]\ar@{-}[ul]\\
 &  & *+[o][F-]{}\ar@{-}[ur]\ar@{-}[ul]
}
\]
\protect\caption{A product of two chains.\label{fig:2dim}}
\end{figure}

\begin{figure}[b]
\[
\xymatrix{ &  &  &  & *+[o][F-]{}\\
 &  & *+[o][F-]{}\ar@{-}[urr]\ar@{-}[dr] & *+[o][F-]{}\ar@{-}[ur]\ar@{-}[dr] & *+[o][F-]{}\ar@{-}[u]\ar@{-}[d] & *+[o][F-]{}\ar@{-}[ul]\ar@{-}[dl] & *+[o][F-]{}\ar@{-}[ull]\ar@{-}[drr]\\
*+[o][F-]{}\ar@{-}[urr]\ar@{-}[drr] & *+[o][F-]{}\ar@{-}[ur]\ar@{-}[dr] & *+[o][F-]{}\ar@{-}[u]\ar@{-}[d] & *+[o][F-]{}\ar@{-}[ul]\ar@{-}[dl] & *+[o][F-]{}\ar@{-}[urr]\ar@{-}[ull]\ar@{-}[drr] & *+[o][F-]{}\ar@{-}[ur]\ar@{-}[dr] & *+[o][F-]{}\ar@{-}[u]\ar@{-}[d] & *+[o][F-]{}\ar@{-}[ul]\ar@{-}[dl] & *+[o][F-]{}\\
 &  & *+[o][F-]{}\ar@{-}[ur]\ar@{-}[urr] & *+[o][F-]{}\ar@{-}[ur]\ar@{-}[dr] & *+[o][F-]{}\ar@{-}[u]\ar@{-}[d] & *+[o][F-]{}\ar@{-}[ul]\ar@{-}[dl] & *+[o][F-]{}\ar@{-}[urr]\ar@{-}[ul]\\
 &  &  &  & *+[o][F-]{}\ar@{-}[urr]\ar@{-}[ull]
}
\]
\protect\caption{A planar modular lattice.\label{fig:2dim-1}}
\end{figure}
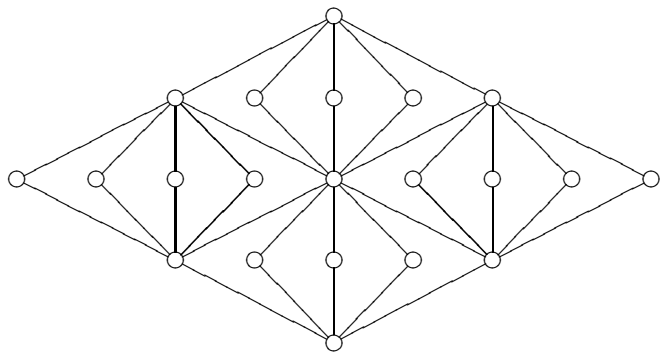

\begin{thm}
\label{thm:distributiv}A distributive lattice is Shannon if and only
if the order dimension at most 3.
\end{thm}
The free distributive lattice with three generators is a lattice on
with the property that any distributive lattice generated by three
elements is isomorphic to a sub-lattice. The free distributive lattice
with three generators has 18 elements \cite[45-46, Theorem 10]{Graetzer1971}
and is three dimensional. Therefore we get the following result.
\begin{cor}
Any distributive lattice with 3 generators is a Shannon lattice.
\end{cor}
With three generators one can also define the free modular lattice.
This lattice has 28 elements \cite[46-47, Theorem 11]{Graetzer1971}
and by explicit calculations one can check that it is a Shannon lattice.
\begin{prop}
The free modular lattice with 3 generators is a Shannon lattice.
\end{prop}
If we do not require that the lattice is modular (or belong to some
other nice lattice variety) the result does not hold. The free lattice
with three elements contain a sub-lattice isomorphic with the four
dimensional Boolean algebra that is not a Shannon lattice. Therefore
it would be interesting to know if there exists larger lattice varieties
that the variety of modular lattices for which a free lattice with
three generators in the variety is a Shannon lattice. 
\begin{thm}
\label{thm:Any-modular-planar}Any modular planar lattice is a Shannon
lattice.
\end{thm}
The proof uses that it has it was recently proved that a planar modular
lattices can be represented as a distributive lattice with a number
of double irreducible elements added \cite{Quackenbush2010} as illustrated
in Figure \ref{fig:2dim-1}. Each of the extreme polymatroid functions
on a planar modular lattice corresponds to a complicated cryptographic
protocol or secrecy sharing scheme. 

\section*{Acknowledgment}

I want to thank S{\o}ren Riis and Sune Jacobsen for useful discussion
during a research stay at Queen Mary University in January 2012.


\pagebreak{}

\appendices{}

This appendix contains two sections with a more careful description
of the relation between functional dependencies, lattices, semi-graphoid
relations and polymatroid functions. The appendix also contains proofs
of some of the theorems stated in the paper. In order to keep within
this note short some proofs have been foreshortened or have been omitted.

\section{The functional dependence lattices}

In this section we shall describe functional dependencies and relate
it to lattice theory. Much of the terminology is taken from database
theory. The relation between functional dependence and lattices has
previously been studied \cite{Demetrovics1989,Demetrovics1992,Levene1995}.

First we shall consider a set of attributes/variables $V_{i}$ and
subsets of this set of variables. Each attribute $V_{i}$ takes values
in some set $W_{i}.$ The set of subsets is also called the power
set and is ordered by inclusion. With this ordering the power set
is a lattice with intersection and union as lattice operations. We
note that the smallest element in the lattice is the empty set $\emptyset$
and the largest element is the whole set. One consider a set of tuples
(records) that all share the same attributes. A \emph{relation} $R$
is a set of tuples and an assignment of a value in $W_{i}$ to each
attribute $V_{i}.$ One may think of a relation as a function from
tuples to the product space $\prod V_{i}.$ If $X$ and $Y$ are sets
of attributes we say that $Y$ functionally dependence on $X$ in
the relation $R$ and write $X\to Y$ if $\prod_{i\in X}V_{i}\left(t_{1}\right)=\prod_{i\in X}V_{i}\left(t_{2}\right)$
implies $\prod_{i\in Y}V_{i}\left(t_{1}\right)=\prod_{i\in Y}V_{i}\left(t_{2}\right).$ 

Inspired by Armstrong's theory of relational databases we say that
a relation $\to$ in a lattice $L$ satisfies \emph{Armstrong's axioms}
if it satisfies the following properties.

\textbf{Transitivity} If $X\to Y$ and $Y\to Z$, then $X\to Z.$ 

\textbf{Reflexivity} If $X\geq Y$, then $X\to Y.$

\textbf{Augmentation} If $X\to Y$, then $X\vee Z\to Y\vee Z.$

Functional dependence $\to$ in a database obviously satisfies these
inference rules so as an axiomatic system it is sound. Armstrong proved
that these axioms form a complete set of inference rules. That means
that if a set $A$ of functional dependencies is given and a certain
functional dependence $x\to y$ holds in any database where all the
functional dependencies in $A$ hold then $x\to y$ holds in that
database. Therefore for any functional dependence $x\to y$ that cannot
be deduced using Armstrong's axioms the exist a database where the
functional dependence is violated \cite{Ullman1989,Levene1999}. As
a consequence there exists a database where a functional dependence
holds if and only if it can be deduced from Armstrong's axioms.
\begin{thm}
A relation $\to$ on the elements of a lattice satisfies Armstrong's
axioms if and only if $\to$ is a preordering that satisfies the following
two properties.

\textbf{Decomposition} If $Z\to X\vee Y$, then $Z\to X$ and $Z\to Y.$

\textbf{Union} If $Z\to X$ and $Z\to Y$, then $Z\to X\vee Y.$\end{thm}
\begin{IEEEproof}
Assume that $\to$ satisfies Armstrong's axioms. Then $X\geq X$ implies
$X\to X$ so that $\to$ is reflexive. To prove the union property
assume that $Z\to X$ and $Z\to Y$. Then $Z\vee Z\to X\vee Z$ and
$X\vee Z\to X\vee Y$ by augmentation. Then transitivity implies $Z\to X\vee Y.$
To prove the decomposition property assume that $Z\to X\vee Y$. In
the lattice we have $X\vee Y\geq X$ and by reflexivity $X\vee Y\to X$.
Now transitivity implies $Z\to X$. In the same way it is proved that
$Z\to Y.$

Assume that $\to$ is a preordering that satisfies decomposition and
union. To prove reflexivity assume that $X\geq Y.$ Then $X\to X\vee Y$,
which according to the decomposition property implies $X\to Y.$ To
prove augmentation assume that $X\to Y$. We have $X\vee Z\to X$
which together with transitivity implies $X\vee Z\to Y.$ By reflexivity
we have $X\vee Z\to Z.$Therefore the union property implies that
then $X\vee Z\to Y\vee Z.$
\end{IEEEproof}
The first half of this proof was essentially given by Armstrong.

Let $L$ denote a lattice with a relation $\to$ such that Armstrong's
axioms are satisfied. For simplicity assume that $L$ is finite. For
$X\in L$ define $cl\left(X\right)$ as $\bigvee Y_{i}$ where the
join is taken over all $Y_{i}$ such that $X\to Y_{i}.$ The union
property implies that $cl\left(X\right)$ is maximal in the set of
variables determined by $X$. With these definitions we see that $X\to Y$
if and only if $cl\left(X\right)\geq cl\left(Y\right).$ For a relation
that satisfies Armstrong's axioms the unary operator $cl$ satisfies
the following conditions:

\textbf{Extensivity} $X\leq cl\left(X\right).$

\textbf{Isotony} $X\leq Y$ implies $cl\left(X\right)\leq cl\left(Y\right).$

\textbf{Idempotens} $cl\left(X\right)=cl\left(cl\left(X\right)\right).$

An unary operator that satisfies these three properties is called
a \emph{closure operator}. We say that $X$ is closed if $cl\left(X\right)=X.$
If $X$ and $Y$ are closed for some closure operator $cl$ then it
is easy to prove that $X\wedge Y$ is closed \cite[Lemma 28]{Gratzer2003}.
A subset of a lattice that is closed under the meet operation, is
called a semi-lattice or a \emph{closure system.} In \cite{Caspard2003}
closure systems were studied in more detail in the case where the
lattice is a power set. The elements of the closure system are closed
elements under the closure operator defined by $cl\left(\ell\right)=\bigwedge_{x\geq\ell,x\in A}x.$ 
\begin{prop}
Let $\left(L,\leq\right)$ denote a finite lattice. Assume that a
subset $A$ of $L$ is closed under the meet operation. Then $A$
is a lattice under the ordering $\leq.$\end{prop}
\begin{IEEEproof}
The set $A$ is partially ordered by $\leq$ so we just have to prove
that any pair of elements in $A$ has a least upper bound and a greatest
lower bound. The greatest lower bound of $x,y\in A$ is $x\wedge y.$
The least upper bound of $x$ and $y$ is $\bigwedge_{x\vee y\leq z,z\in A}z.$ 
\end{IEEEproof}
The lattice operations in $A$ are given by $X\wedge_{A}Y=X\wedge Y$
and $X\vee_{A}Y=cl\left(X\vee Y\right).$ In particular the closed
elements of a functional dependence relation form a lattice, and this
was essentially the main result of Armstrong although he did not use
lattice terminology. The converse of Armstrong's results is also true: 
\begin{thm}
Let $\left(L,\leq\right)$ denote a finite lattice with a closure
system $A.$ Then the relation $x\to y$ is defined by $cl\left(x\right)\geq cl\left(y\right)$
satisfies Armstrong's axioms.\end{thm}
\begin{IEEEproof}
It is easy to see that $\to$ defines a preordering. The union property
is proved as follows. Assume that $x\to y$ and $x\to z.$ Then $\mbox{cl\ensuremath{\left(x\right)\geq}cl\ensuremath{\left(y\right)\geq}y and \ensuremath{cl\left(x\right)\geq cl\left(z\right)\geq z.}Hence \ensuremath{cl\left(x\right)\geq y\vee z}}$
and $cl\left(x\right)\geq cl\left(y\vee z\right)$ so that $x\to y\vee z.$
The decomposition property is proved by reversing the argument.
\end{IEEEproof}
The theorem as it is formulated here probably appear somewhere in
the literature on lattices although the author has not been able to
locate a good reference.
\begin{example}
We consider three variables $a,b$ and $c$ that denote real numbers.
Assume that $c=\left(a+b\right)^{2}.$ Then the associated lattice
is the lattice that is normally called $S_{7}.$ This is illustrated
in Figure \ref{fig:S7}.

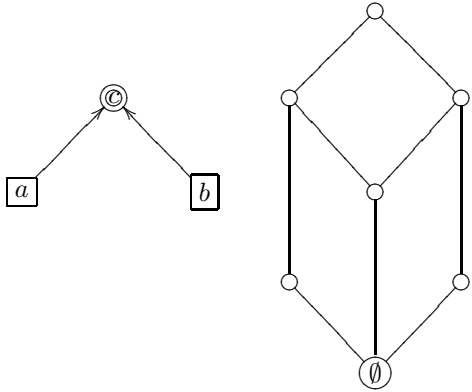
\begin{figure}[tbh]
\[
\xymatrix{ &  &  &  & *+[o][F]{}\\
 & *+[o][F=]{c} &  & *+[o][F]{}\ar@{-}[ur]\ar@{-}[dr] &  & *+[o][F]{}\ar@{-}[ul]\ar@{-}[dl]\\
*+[F]{a}\ar[ur] &  & *+[F]{b}\ar[ul] &  & *+[o][F]{}\ar@{-}[dd]\\
 &  &  & *+[o][F]{}\ar@{-}[uu] &  & *+[o][F]{}\ar@{-}[uu]\\
 &  &  &  & *+[o][F]{\emptyset}\ar@{-}[ur]\ar@{-}[ul]
}
\]
\protect\caption{To the left and influence diagram for three variables is drawn with
arrows indicating the direction of influence. To the right the Hasse
diagram for the corresponding lattice of functional dependence is
drawn with the smallest element $\left(\emptyset\right)$ indicated.
The name of this lattice is $S_{7}$. \label{fig:S7}}
\end{figure}

\end{example}
Even simple examples of functional dependence lattices may be complicated
to describe if they are not based on simple causal relations between
the variables.
\begin{example}
This example concern fruit from a supermarket. Variable $X$ tells
whether the supermarket will sell it at normal price, or at a reduced
price because it is close to the expiration date, or whether it is
through out because the expiration date has been exceeded. Variable
$Z$ describes whether the fruit tastes very fresh, is eatable, or
looks disgusting. The variable $Y$ tells whether the fruit will make
you sick or not. The functional dependencies are given by $Z\subseteq Y$
and $X\uplus Y=X\uplus Z.$ The lattice is $N_{5}.$ This is the standard
example of a lattice that is not modular.

\begin{figure}[ptb]
\centering{}
\[
\xymatrix{ & *+[o][F]{}\\
 &  & *+[o][F]{z}\ar@{-}[ul]\\
*+[o][F]{x}\ar@{-}[uur]\\
 &  & *+[o][F]{y}\ar@{-}[uu]\\
 & *+[o][F]{\emptyset}\ar@{-}[ur]\ar@{-}[uul]
}
\]
\protect\caption{\label{fig:N5}Hasse diagram of the lattice $N_{5}.$ }
\end{figure}
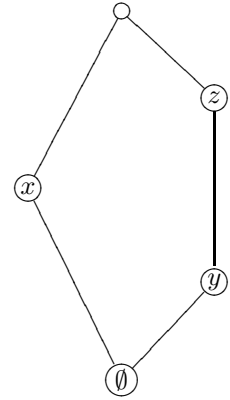
\end{example}
\begin{thm}
Any finite lattice can be represented as a closure system on a power
set\end{thm}
\begin{IEEEproof}
Let $L$ be a lattice. For each $a\in L$ the principle ideal of $a$
is $\downarrow\left(a\right)=\left\{ x\in L\mid x\leq a\right\} $.
This gives an embedding of $L$ into the power set of $L$ in such
a way that meet in the lattice corresponds to intersection in the
power set. 
\end{IEEEproof}
As a result any lattice is equivalent to a lattice of functional dependence,
so all what can be said about functional dependence can be expressed
in the language of lattices. Most of the time we will formulate our
results in terms of closure systems. Since the notation for inclusion
and intersection is fixed, we will use $\supseteq$ to denote the
ordering of a functional dependence lattice and $\cap$ to denote
the meet operation. If the lattice is the whole power set, i.e. a
Boolean lattice then we will use $\cup$ to denote the join operation.
If we have not assumed that the lattice is Boolean we may use $\vee$
or $\uplus$ or $\sqcup$ or some similar symbol to denote the join
operation.

With the above results we can prove that Armstrong's axioms form a
complete set of inference rules for functional dependencies.
\begin{thm}
\label{thm:completeArmstrong-1}For any finite lattice there exist
a set of related variables such that the elements of the lattice corresponds
to closed sets under functional dependence. \end{thm}
\begin{IEEEproof}
A lattice can be represented as a closure system of a power set of
some set $I.$ To each element $i\in I$ we associate a binary variable
$V_{i}$ with values in $W_{i}=\left\{ 0,1\right\} .$ Let $C$ denote
the closed sets in the power sets. For each $c\in C$ we assign an
tuple $t_{c}$ so that 
\[
V_{i}\left(t_{c}\right)=1\,\textrm{if }i\in c.
\]
Assume that $a\supseteq b.$ If $\left(V_{i}\left(t_{c_{1}}\right)\right)_{i\in a}=\left(V_{i}\left(t_{c_{2}}\right)\right)_{i\in a}$
for all tuples $t_{c_{j}}$ then $\left(V_{i}\left(t_{c_{1}}\right)\right)_{i\in b}=\left(V_{i}\left(t_{c_{2}}\right)\right)_{i\in b}$
for holds for all tuples $t_{c_{j}}$. Hence $a\to b.$ 

Assume that $a\to b.$ According to the definition it means that if
$\left(V_{i}\left(t_{c_{1}}\right)\right)_{i\in a}=\left(V_{i}\left(t_{c_{2}}\right)\right)_{i\in a}$
for all tuples $t_{c_{j}}$ then $\left(V_{i}\left(t_{c_{1}}\right)\right)_{i\in b}=\left(V_{i}\left(t_{c_{2}}\right)\right)_{i\in b}$
for holds for all tuples $t_{c_{j}}$. Assume that $\left(V_{i}\left(t_{c_{1}}\right)\right)_{i\in a}=\left(V_{i}\left(t_{c_{2}}\right)\right)_{i\in a}.$
Then for all $i\in a$ we have $V_{i}\left(t_{c_{1}}\right)=V_{i}\left(t_{c_{2}}\right)$
which is equivalent to $a\cap c_{1}=a\cap c_{2}.$ Similarly $\left(V_{i}\left(t_{c_{1}}\right)\right)_{i\in b}=\left(V_{i}\left(t_{c_{2}}\right)\right)_{i\in b}$
is equivalent to $b\cap c_{1}=b\cap c_{2}.$ Choose $c_{1}=a$ and
$c_{2}=a\uplus b$. Then $a\cap c_{1}=a\cap c_{2}$ is automatically
fulfilled and $b\cap c_{1}=b\cap c_{2}$ can be rewritten as $b\cap a=b\cap\left(a\uplus b\right)=b$,
which implies that \textbf{$a\supseteq b.$}
\end{IEEEproof}

\section{Independence in lattices}

In statistics one studies the relation $\left(A\bot B\mid C\right)$
meaning that $A$ and $B$ are independent given $\check{s}C$, where
$A,$ $B$ and $C$ are disjoint subsets of a set $M$ of random variables
with respect to a probability measure. We will call this notion of
independence \emph{statistical independence}. 

We shall say that a relation $\left(\cdot\bot\cdot\mid\cdot\right)$
on a lattice $\left(L,\cap,\uplus\right)$ is a \emph{semi-graphoid
relation}, if it satisfies the following axioms:

\textbf{Existence} $\left(X\bot Y\mid X\right).$

\textbf{Symmetry} $\left(X\bot Y\mid W\right)$ if and only if $\left(Y\bot X\mid W\right).$

\textbf{Decomposition} If $\left(X\bot Y\uplus Z\mid W\right)$ then
$\left(X\bot Z\mid W\right).$

\textbf{Contraction} $\left(X\bot Z\mid W\right)$ and $\left(X\bot Y\mid Z\uplus W\right)$
implies $\left(X\bot Y\uplus Z\mid W\right).$ 

\textbf{Weak union} $\left(X\bot Y\uplus Z\mid W\right)$ implies
$\left(X\bot Y\mid Z\uplus W\right).$

These properties should hold for all $X,Y,Z,W\in L.$ We note that
statistical independence with respect to a probability measure is
semi-graphoid. In this paper we are particularly interested in the
case where the subsets are not disjoint. A relation that satisfies
the last for properties for disjoint sets in a power was said to be
semi-graphoid \cite{Pearl88}. In a recent paper \cite{Paolini2014}
a much longer list of axioms for the notion of independence was given.
Most of those axioms can be proved from the axioms stated in this
paper.
\begin{thm}
A semi-graphoid relation $\left(\cdot\bot\cdot\mid\cdot\right)$ satisfies
the following properties.

\textbf{Reflexivity} For all $A$ we have $\left(X\bot X\mid X\right).$

\textbf{Normality} If $\left(X\bot Y\mid W\right)$ then $\left(X\bot Y\uplus W\mid W\right).$

\textbf{Monotonicity} If $\left(X\bot Y\mid W\right)$ and $Y\uplus W\supseteq Z$
then $\left(X\bot Z\mid W\right).$

\textbf{Triviality} $\left(X\bot\emptyset\mid Y\right)$

\textbf{Base monotonicity} If $\left(A\bot B\mid D\right)$ and $B\supseteq C\supseteq D$
then $\left(A\bot B\mid C\right).$

\textbf{Transitivity} If $\left(A\bot B\mid C\right)$ and $\left(A\bot C\mid D\right)$
and $B\supseteq C\supseteq D$ then $\left(A\bot B\mid D\right).$

\textbf{Autonomy} If $\left(A\bot A\mid C\right)$ then $\left(A\bot B\mid C\right).$
\end{thm}
In a power set of random variables we note that if $A$ is independent
of $A$ given $C$ then $A$ is a function of $C$ almost surely.
If $\left(B\bot B\mid A\right)$ we write $A\supseteq_{\bot}B$. 
\begin{defn}
An semi-graphoid relation is said to be consistent with $\subseteq$
if $X\subseteq Y$ is equivalent to $\left(X\bot X\mid Y\right)$.\end{defn}
\begin{thm}
If $\left(L,\cap,\uplus\right)$ is a lattice with a semi-graphoid
relation $\left(\cdot\bot\cdot\mid\cdot\right)$ then the relation
$\supseteq_{\bot}$ satisfies Armstrong's axioms. The relation $\left(\cdot\bot\cdot\mid\cdot\right)$
restricted to the lattice of closed lattice elements is semi-graphoid. \end{thm}
\begin{IEEEproof}
\textbf{Reflexivity of $\supseteq_{\bot}$} This follows according
to the reflexivity property of $\bot.$

\textbf{Transitivity} Assume that $X\supseteq_{\bot}Y$ and $Y\supseteq_{\bot}Z$.
Autonomy implies that $\left(Z\bot Z\uplus X\mid Y\right)$ and by
weak union $\left(Z\bot Z\mid Y\uplus X\right).$ Autonomy and $X\supseteq_{\bot}Y$
together imply that $\left(Y\bot Z\mid X\right).$ Contraction then
implies $\left(Z\bot Y\uplus Z\mid X\right)$. Decomposition gives
$\left(Z\bot Z\mid X\right).$ 

\textbf{Decomposition} This follows from the decomposition property
of $\bot.$

\textbf{Union} Assume that $X\supseteq_{\bot}Y$ and $X\supseteq_{\bot}Z$.
Then $\left(Y\bot Y\mid X\right)$ and $\left(Z\bot Z\mid X\right)$
and by autonomy $\left(Y\bot Y\uplus Z\mid X\right)$ and $\left(Z\bot Y\uplus Z\uplus Y\mid X\right).$
Hence $\left(Z\bot Y\uplus Z\mid Y\uplus X\right)$ by weak union
and $\left(Y\uplus Z\bot Y\uplus Z\mid X\right)$ by contraction.

For the last result one just has to prove that $\left(X\bot Y\mid Z\right)$
if and only if $\left(X\bot cl_{\bot}\left(Y\right)\mid Z\right)$
if and only if $\left(X\bot Y\mid cl_{\bot}\left(Z\right)\right)$.
This follows from Armstrong's results. 
\end{IEEEproof}
The significance of this theorem is that if we start with a semi-graphoid
relation on a lattice then this semi-graphoid relation is also semi-graphoid
when restricted elements that are closed under functional dependence.
\begin{thm}
Any finite lattice can be represented as a closure system of an semi-graphoid
relation defined on a power-set.\end{thm}
\begin{IEEEproof}
For any finite lattice $L$ one identify the elements by sets of binary
variables $v_{i}$, and a relation can be defined where the tuples
have the form $i_{c},c\in L$ as in the proof of Theorem \ref{thm:completeArmstrong-1}.
Each tuple can be identified with a point in the product space $\prod W_{i}.$
Assign a uniform distribution to each point in the product space.
With respect to this probability measure $\left(b\bot b\mid a\right)$
if and only if $a$ determines $b$ almost surely. Since the probability
measure is discrete $\left(b\bot b\mid a\right)$ is valid if and
only if $a\supseteq b.$
\end{IEEEproof}
The semi-graphoid relation defined in the proof of the previous theorem
is based on the uniform distribution on the tuples. We note that any
other distribution that has positive probability on the same tuples
will also give a representation of the lattice in terms of a semi-graphoid
relation. For disjoint sets independence will depend on the choice
of probability distribution.

\section{Proof of Proposition \ref{prop:meetsemilattice}}

Assume that $L$ is a Shannon lattice and that $M$ is a sub-lattice.
Let $h:M\to\mathbb{R}$ denote a polymatroid function. For $\ell\in L$
let $\tilde{\ell}$ denote the $m\in M$ that minimize $h\left(m\right)$
under the constraint that $m\supseteq\ell.$ Define the function $\tilde{h}\left(\ell\right)=h\left(\tilde{\ell}\right).$
Now $\tilde{h}$ is an extension of $h$ and with this definition
$\tilde{h}$ is non-negative and increasing. For $x,y\in L$ we have
\begin{eqnarray*}
\tilde{h}\left(x\right)+\tilde{h}\left(y\right) & = & h\left(\tilde{x}\right)+h\left(\tilde{y}\right)\\
 & \geq & h\left(\tilde{x}\uplus\tilde{y}\right)+h\left(\tilde{x}\cap\tilde{y}\right)\\
 & \geq & \tilde{h}\left(x\uplus y\right)+\tilde{h}\left(x\cap y\right)
\end{eqnarray*}
because $\tilde{x}\uplus\tilde{y}\geq x\uplus y$ and $\tilde{x}\cap\tilde{y}\geq x\cap y.$
Hence $\tilde{h}$ is submodular. By the assumption $\tilde{h}$ is
entropic so the restriction of $\tilde{h}$ to $M$ is also entropic.

\section{Proof of Lemma \ref{lem:increasing}}

Assume that $h$ is submodular and increasing on $\cap$-irreducible
elements. We have to prove that if $a\supseteq c$ then $h\left(a\right)\geq h\left(c\right).$
In order to obtain a contradiction assume that $c$ is a maximal element
such that there exist an element $a$ such $a\supseteq c$ but $h\left(a\right)<h\left(c\right).$
We may assume that $a$ cover $c$. Since $h$ is increasing at $\cap$-irreducible
elements $c$ cannot be $\cap$-ireducible. Therefore there exists
a maximal element $b$ such that $b\supseteq c$ but $b\nsupseteq a.$
Since $a$ cover $c$ we have $a\cap b=c.$ According to the assumptions
$h\left(a\right)+h\left(b\right)\geq h\left(a\uplus b\right)+h\left(a\cap b\right)$
and $h\left(a\uplus b\right)\geq h\left(b\right)$ because $c$ is
a maximal element that violates monotonicity. Therefore $h\left(a\right)\geq h\left(a\cap b\right)=h\left(c\right).$

\section{Proof of Theorem \ref{thm:M_n}}

Let the values in the double irreducible elements be denoted $x_{1},x_{2},\dots,x_{n-2}$.
If $n=1$ the extreme polymatroid functions are $x_{1}=0$ and $x_{1}=1$
and these points are obviously entropic. If $n=4$ the extreme points
are $\left(x_{1},x_{2}\right)=\left(0,1\right)$ and $\left(x_{1},x_{x}\right)=\left(1,0\right)$
and $\left(x_{1},x_{2}\right)=\left(1,1\right),$which are all entropic.

Assume $n\geq5$. Then the values should satisfy the inequalities
\begin{eqnarray*}
0\leq & x_{i} & \leq1\\
x_{i}+x_{j} & \geq & 1.
\end{eqnarray*}
If $\left(x_{1},x_{2},\dots,x_{n-2}\right)$ is an extreme point then
each variables should satisfy one of the inequalities with equality.
Assume $x_{i}=0.$ Then sub-modularity implies that $x_{j}=1$ for
$j\neq i$. The extreme point $\left(0,0,\dots,0,1,0,\dots,0\right)$
is obviously entropic. If $x_{i}=1$ this gives no further constraint
on the other values, so it corresponds to an extreme point on a lattice
with one less variable. Finally assume that $x_{i}+x_{j}=1$ for all
$i,j$. Then $x_{i}=\nicefrac{1}{2}$ for all $i.$ 

We have to find $n-2$ random variables $X_{1},X_{2}\dots,X_{n-2}$
that are independent but such that any two determine the rest. Let
$p$ denote a prime larger than $n-2$. Let $Y$ and $Z$ denote independent
random variables with values in $\mathbb{Z}_{p}$ each with a uniform
distribution. If $X_{j}$ is defined to be equal to $Y+jZ$ then the
variables $X_{j}$ are mutually independent and any pair of these
random variables determine all the other variables. 

Instead of constructing the variables $X_{1},X_{2},\dots,X_{n-2}$
we can find a group $G$ and subgroups $G_{1},G_{2},\dots,G_{n-2}$such
that $\left|G\right|=2\left|G_{i}\right|$ using general results about
entropy inequalities and groups \cite{Chan2002}. The group $G$ can
be chosen as $\mathbb{Z}_{p}\times\mathbb{Z}_{p}$where $p$ is some
prime number greater than $n-2.$ The group $G$ has $p+1$ subgroups
isomorphic to $\mathbb{Z}_{p}.$

\section{Proof of Lemma \ref{lem:Modulaer}}

Let $a$ and $b$ denote two lattice elements. We have to prove that
$h\left(a\right)+h\left(b\right)\geq h\left(a\uplus b\right)+h\left(a\cap b\right).$ 

Assume that $x_{1},x_{2},\dots,x_{n}$ is sequence of elements such
$a\cap b\subseteq x\subseteq x_{2}\subseteq\dots\subseteq x_{n}\subseteq a.$
Define $y_{i}=x_{i}\uplus b.$ Then modularity implies 
\begin{eqnarray*}
x_{i+1}\cap y_{i} & = & x_{i+1}\cap\left(b\uplus x_{i}\right)\\
 & = & \left(x_{i}\cap b\right)\uplus\left(x_{i+1}\cap x_{i}\right)\\
 & = & \left(a\cap b\right)\uplus x_{i}\\
 & = & x_{i}.
\end{eqnarray*}
 We also have 
\begin{eqnarray*}
x_{i+1}\uplus y_{i} & = & x_{i+1}\uplus\left(b\uplus x_{i}\right)\\
 & = & \left(x_{i+1}\uplus x_{i}\right)\uplus b\\
 & = & x_{i+1}\uplus b\\
 & = & y_{i+1}.
\end{eqnarray*}
Assume that the modular inequality holds for all the sub-lattices
$L_{i}=\left\{ x_{i},x_{i+1},y_{i},y_{i+1}\right\} .$ Then we can
add all the inequalities $h\left(x_{i+1}\right)+h\left(y_{i}\right)\leq h\left(x_{i}\right)+h\left(y_{i+1}\right)$
to get $h\left(x_{n}\right)+h\left(y_{1}\right)\leq h\left(x_{1}\right)+h\left(y_{n}\right).$
Note that we can choose the sequence $x_{1},x_{2},\dots,x_{n}$ such
that $x_{i+1}$ covers $x_{i}$ and such that $x_{1}=a\cap b$ and
$x_{n}=a.$ Therefore we it is sufficient to prove that $h\left(a\right)+h\left(b\right)\geq h\left(a\uplus b\right)+h\left(a\cap b\right)$
when $a$ cover $a\cap b.$

Similarly it is sufficient to prove that $h\left(a\right)+h\left(b\right)\geq h\left(a\uplus b\right)+h\left(a\cap b\right)$
when $b$ cover $a\cap b.$ If $a$ and $b$ both cover $a\cap b$
then $M=\left\{ a,b,a\cap b,a\uplus b\right\} $ is a sub-lattice
of $x^{+}$ if $x=a\cap b.$

\section{Proof of Theorem \ref{thm:distributiv}}

If the lattice is one-dimensional we just get a deterministic Markov
chain for which positivity and monotonicity are sufficient conditions
for a function to be entropic. Assume that the lattice is two-dimensional. 

We will show that an extreme polymatroid function only takes the values
0 and 1. Assume that $\left(A,\leq\right)$ the poset of irreducible
elements in the distributive lattice. The proof is by induction on
the number of elements $k$ in the lattice. If $k=2$ this is obvious.
Assume that it has been proved for all distributive lattices where
$k\leq n$ and let $L$ be a lattice with $n+1$ elements. We note
that a distributive lattice is modular. Therefore it is sufficient
that the sub-modular inequality is satisfied on sub-lattices of the
form $x^{+}.$ We know that he lattice is sub-lattice of a product
of two totally ordered lattices. Such a product lattice is planer
in the sense that it has a Hasse diagram without intersection lines.
The Hasse diagram consists of small squares each representing a sub-lattice
of the form $x^{+}$. 

Now consider a polymatroid  function $h$ on the lattice. Assume that
$h$ is an extreme point in the set of all polymatroid functions.
For each point in the lattice the value is constrained by a number
of inequalities and since we have assumed that the function is an
extreme point at least one of the inequalities holds with equality.
We start at the double irreducible element $b$. It is contained by
two monotone inequalities and one submodular inequality.
\begin{eqnarray}
h\left(a\right) & \leq & h\left(b\right)\nonumber \\
h\left(b\right) & \leq & h\left(d\right)\nonumber \\
h\left(a\right)+h\left(d\right) & \leq & h\left(b\right)+h\left(c\right).\label{eq:betingelser}
\end{eqnarray}
The submodular inequality implies that $h\left(b\right)\geq h\left(a\right)+\left(h\left(d\right)-h\left(c\right)\right)$
which a stronger condition than the first monotone condition. Therefore
the conditions on $y_{1}$are
\begin{equation}
h\left(a\right)+h\left(d\right)-h\left(c\right)\leq h\left(b\right)\leq h\left(d\right).\label{eq:betingelser-1}
\end{equation}
Observe that $h\left(a\right)+h\left(d\right)-h\left(c\right)\leq h\left(d\right)$.
Since both the lower bound on $h\left(b\right)$ and the upper bound
on $h\left(b\right)$ are linear any extreme polymatroid is also an
extreme polymatroid when it is restricted to the lattice where the
element $b$ has been removed. According to the induction hypothesis
such an extreme polymatroid function only takes the values 0 and 1.
Therefore if the polymatroid function on the original lattice is extreme
one of the inequalities in (\ref{eq:betingelser-1}) must hold with
equality and therefore $h\left(b\right)=0$ or $h\left(b\right)=1.$
is entropic.
\begin{figure}[b]
\[
\xymatrix{ &  &  &  & *+[o][F-]{22}\\
 &  & *+[o][F-]{21}\ar@{-}[urr]\ar@{-}[dr] & *+[o][F-]{111}\ar@{-}[ur]\ar@{-}[dr] & *+[o][F-]{112}\ar@{-}[u]\ar@{-}[d] & *+[o][F-]{113}\ar@{-}[ul]\ar@{-}[dl] & *+[o][F-]{12}\ar@{-}[ull]\ar@{-}[drr] &  & ]\\
*+[o][F-]{20}\ar@{-}[urr]\ar@{-}[drr] & *+[o][F-]{101}\ar@{-}[ur]\ar@{-}[dr] & *+[o][F-]{102}\ar@{-}[u]\ar@{-}[d] & *+[o][F-]{103}\ar@{-}[ul]\ar@{-}[dl] & *+[o][F-]{11}\ar@{-}[urr]\ar@{-}[ull]\ar@{-}[drr] & *+[o][F-]{011}\ar@{-}[ur]\ar@{-}[dr] & *+[o][F-]{011}\ar@{-}[u]\ar@{-}[d] & *+[o][F-]{013}\ar@{-}[ul]\ar@{-}[dl] & *+[o][F-]{02}\\
 &  & *+[o][F-]{10}\ar@{-}[ur]\ar@{-}[urr] & *+[o][F-]{001}\ar@{-}[ur]\ar@{-}[dr] & *+[o][F-]{002}\ar@{-}[u]\ar@{-}[d] & *+[o][F-]{003}\ar@{-}[ul]\ar@{-}[dl] & *+[o][F-]{01}\ar@{-}[urr]\ar@{-}[ul]\\
 &  &  &  & *+[o][F-]{00}\ar@{-}[urr]\ar@{-}[ull]
}
\]
\protect\caption{A planar modular lattice with indexing of the elements.\label{fig:2dim-1-1}}
\end{figure}
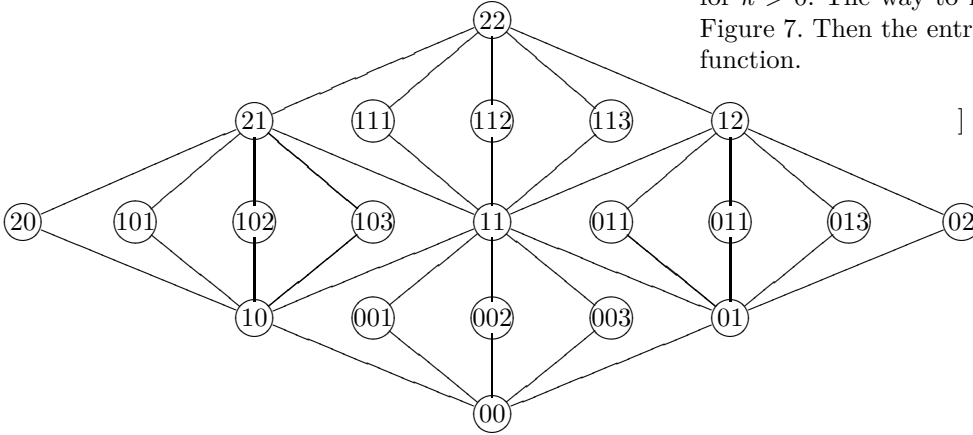

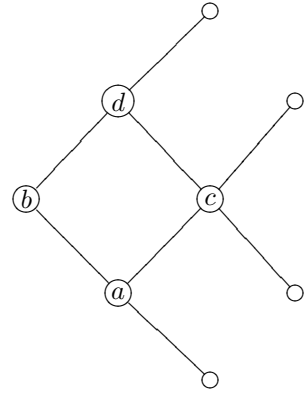
\begin{figure}[tbh]
\[
\xymatrix{ &  & *+[o][F-]{}\\
 & *+[o][F-]{d}\ar@{-}[ur]\ar@{-}[dl] &  & *+[o][F-]{}\\
*+[o][F-]{b} &  & *+[o][F-]{c}\ar@{-}[ur]\ar@{-}[ul]\\
 & *+[o][F-]{a}\ar@{-}[ur]\ar@{-}[ul] &  & *+[o][F-]{}\ar@{-}[ul]\\
 &  & *+[o][F-]{}\ar@{-}[ul]
}
\]
\protect\caption{The upper right corner of the lattice.}
\end{figure}

Since an extreme polymatroid function only takes the values 0 and
1 the lattice generated by the polymatroid function has only two elements
and this is obviously entropic.

If the lattice is three dimensional one have to modify the above procedure.
A three dimensional distributive lattice may not have any double irreducible
elements. If a single element is deleted from the lattice it is no
longer modular, but modularity is needed if we should use Lemma \ref{lem:Modulaer}.
Instead one consider sequences $\left(a_{1},a_{2}\dots a_{n}\right),\left(b_{1},b_{2}\dots b_{n}\right),\left(c_{1},c_{2},\dots c_{n}\right),$
and $\left(d_{1},d_{2}\dots d_{n}\right)$ with the conditions
\begin{alignat*}{1}
h\left(a_{j}\right)+h\left(d_{j}\right)-h\left(c_{j}\right) & \leq h\left(b_{j}\right)\leq h\left(d_{j}\right)\\
h\left(d_{j+1}\right)-h\left(d_{j}\right)\leq h\left(b_{j+1}\right)- & h\left(b_{j}\right)\leq h\left(b_{j+1}\right)-h\left(b_{j}\right).
\end{alignat*}
One can then prove that any extreme polymatroid function only takes
the values 0,$\nicefrac{1}{2}$, and 1 by a more complicated induction
argument. One can then use \ref{cor:trevaerdier} to include that
any extreme polymatroid function

\section{Proof of Theorem \ref{thm:Any-modular-planar}}

We use that it has it was recently proved that a planar modular lattices
can be represented as a distributive lattice with a number of double
irreducible elements added \cite{Quackenbush2010}. The proof has
the same structure as for distributive lattices, but the existence
of the double irreducible elements implies that there are also extreme
polymatroid functions that are proportional to the ranking function.
Let $X_{1},X_{2},\dots X_{m},Y_{1},Y_{2},\dots,Y_{n}$ denote independent
random variables uniformly distributed over $\mathbb{Z}_{p}$ for
some large value of $p.$ Let $Z_{ij}$ denote the random variable
\[
\biguplus_{\ell\leq i}X_{\ell}\uplus\biguplus_{\ell\leq j}Y_{\ell}.
\]
and let $Z_{ijk}$ denote the random variable 
\[
\biguplus_{\ell\leq i}X_{\ell}\uplus\biguplus_{\ell\leq j}Y_{\ell}\uplus\left(X_{i+1}+k\cdot Y_{j+1}\right)
\]
for $k>0.$ The way to index the variables can be seen in Figure \ref{fig:2dim-1-1}.
Then the entropy is proportional to the ranking function.

\begin{thebibliography}{10}
\providecommand{\url}[1]{#1}
\csname url@samestyle\endcsname
\providecommand{\newblock}{\relax}
\providecommand{\bibinfo}[2]{#2}
\providecommand{\BIBentrySTDinterwordspacing}{\spaceskip=0pt\relax}
\providecommand{\BIBentryALTinterwordstretchfactor}{4}
\providecommand{\BIBentryALTinterwordspacing}{\spaceskip=\fontdimen2\font plus
\BIBentryALTinterwordstretchfactor\fontdimen3\font minus
  \fontdimen4\font\relax}
\providecommand{\BIBforeignlanguage}[2]{{%
\expandafter\ifx\csname l@#1\endcsname\relax
\typeout{** WARNING: IEEEtran.bst: No hyphenation pattern has been}%
\typeout{** loaded for the language `#1'. Using the pattern for}%
\typeout{** the default language instead.}%
\else
\language=\csname l@#1\endcsname
\fi
#2}}
\providecommand{\BIBdecl}{\relax}
\BIBdecl

\bibitem{Zhang1988}
Z.~Zhang and R.~W. Yeung, ``On characterization of entropy function via
  information inequalities,'' \emph{IEEE Trans . Inform. Theory}, vol.~44,
  no.~4, pp. 1440--1452, July 1988.

\bibitem{Matus2007}
F.~Mat{\'u}s, ``Infinitely many information inequalities,'' in \emph{Proc.
  International Symposium on Information Theory (ISIT) 2007. Nice France}, June
  2007, pp. 2101--2105.

\bibitem{Demetrovics1989}
J.~Demetrovics, L.~Libkin, and I.~B. Muchnik, ``Functional dependencies and the
  semilattice of closed classes,'' in \emph{MFDBS '89 Proceedings of the 2nd
  Symposium on Mathematical Fundamentals of Database Systems}.\hskip 1em plus
  0.5em minus 0.4em\relax Springer, 1989, pp. 136--147.

\bibitem{Demetrovics1992}
------, ``Functional dependencies in relational databases: A lattice point of
  view,'' \emph{Discrete Applied Mathematics}, vol.~40, no.~2, pp. 155--185,
  Dec. 1992.

\bibitem{Levene1995}
M.~Levene, ``A lattice view of functional dependencies in incomplete
  relations,'' \emph{Acta Cybernetica}, vol.~12, pp. 181--207, 1995.

\bibitem{Harremoes2011g}
\BIBentryALTinterwordspacing
P.~Harremo{\"e}s, ``Functional dependences and {B}ayesian networks,'' in
  \emph{Proceedings WITMSE 2011}, Helsinki, 2011. [Online]. Available:
  \url{http://www.harremoes.dk/Peter/FunctionalWITSME.pdf}
\BIBentrySTDinterwordspacing

\bibitem{Harremoes2015}
\BIBentryALTinterwordspacing
------, ``Influence diagrams as convex geometries,'' 2015, submitted. [Online].
  Available: \url{www.harremoes.dk/Peter/FunctionalDAG.pdf}
\BIBentrySTDinterwordspacing

\bibitem{Armstrong1974}
W.~W. Armstrong, ``Dependency structures of data base relationships,'' in
  \emph{IFIP Congress}, 1974, pp. 580--583.

\bibitem{Ullman1989}
J.~D. Ullman, \emph{Principles of Database and Knowledge-base Systems}.\hskip
  1em plus 0.5em minus 0.4em\relax Stanford: Computer Science Press, 1989,
  vol.~1.

\bibitem{Levene1999}
M.~Levene and G.~Loizou, \emph{A Guide Tour of Relational Databases and
  Beyond}.\hskip 1em plus 0.5em minus 0.4em\relax Springer, 1999.

\bibitem{Schmidt1994}
R.~Schmidt, \emph{Subgroup Lattices of Groups}.\hskip 1em plus 0.5em minus
  0.4em\relax Walter de Gruyter, 1994.

\bibitem{Behrendt1991}
G.~Behrendt, ``Representations of locally distributive lattice,''
  \emph{Portugaliae Mathematica}, vol.~48, no.~3, pp. 351--355, 1991.

\bibitem{Guille2011}
L.~Guille, T.~Chan, and A.~Grant, ``The minimal set of {I}ngleton
  inequalities,'' \emph{IEEE Trans. Inform. Theory}, vol.~57, no.~4, pp.
  1849--1864, April 2011.

\bibitem{Niepert2013}
M.~Niepert, M.~Gyssens, B.~Sayrafi, and D.~V. Gucht, ``On the conditional
  independence implictation problem: A lattice-theoretic approach,''
  \emph{Artificial Intelligence}, vol. 202, pp. 29--51, 2013.

\bibitem{Dilworth1950}
R.~P. Dilworth, ``A decomposition theorem for partially ordered sets,''
  \emph{Ann. of Math.}, vol.~51, no.~2, pp. 161--166, 1950.

\bibitem{Graetzer1971}
G.~Gr{\"a}tzer, \emph{Lattice Theory}.\hskip 1em plus 0.5em minus 0.4em\relax
  Dover, 1971.

\bibitem{Quackenbush2010}
G.~G.~W. Quackenbush, ``The variety generated by planar modular lattices,''
  \emph{Algebra universalis}, vol.~63, no. 2-3, pp. 187--201, 2010.

\bibitem{Gratzer2003}
G.~Gr{\"a}tzer, \emph{General Lattice Theory}, second edition~ed.\hskip 1em
  plus 0.5em minus 0.4em\relax Birkh{\"a}user, 2003.

\bibitem{Caspard2003}
N.~Caspard and B.~Monjardet, ``The lattices of closure systems, closure
  operators, and implicational systems on a finite set: a survey,''
  \emph{Discrete Applied Mathematics}, vol. 127, no.~2, pp. 241--269, 2003.

\bibitem{Paolini2014}
\BIBentryALTinterwordspacing
G.~Paolini, ``Independence logic and abstract independence relations,'' Sept.
  2014, to appear in Mathematical Logic Quarterly. [Online]. Available:
  \url{http://arxiv.org/pdf/1401.6907.pdf}
\BIBentrySTDinterwordspacing

\bibitem{Chan2002}
T.~H. Chan and R.~W. Yeung, ``On a relation between information inequalities
  and group theory,'' \emph{IEEE Trans. Inform. Theory}, vol.~48, no.~7, pp.
  1992--1995, 2002.

\end{thebibliography}
\end{document}